\documentclass{cimento}

\usepackage{graphicx}
\usepackage{epsfig}

\title{The Role of Core in the Instability of Spherical Gravitating Systems}

\instlist{
\inst{ins:YerPhi} Yerevan Physics Institute, Yerevan, Armenia
}

\PACSes{
\PACSit{05.70.Ln}{Nonequilibrium thermodynamics, irreversible processes}
\PACSit{98.52.-b}{Normal galaxies; extragalactic objects and systems}
}

\author{
T.~Ghahramanyan\from{ins:YerPhi},
}
\begin{document}
\maketitle

\begin{abstract}
We study the dynamical instability in spherical gravitating systems with core using the Ricci curvature criterion. By means of numerical estimations of the Ricci curvature for static N-body systems, it is shown that the core can both, increase and decrease the degree of instability in the system.  This behavior is determined by the radius of the core and differs from the role of a central mass. 
\end{abstract}

\section{The Problem}

Typical stellar systems, globular clusters and elliptical galaxies, possess dense cores which have to influence
the dynamical properties of the systems. At the same time, the gravitating N-body systems as non-linear many dimensional systems, possess chaotic properties \cite{GP,GIAU,Ben}. 
In the present paper we study the role of the core in the chaotic properties of the spherical collisionless gravitating systems. Namely, the question is, whether the core is increasing or decreasing the relative
chaos in the system. For comparison, it is known, that a massive center is increasing the chaotic properties
of the system \cite{GK,Trem,EZ} with respect to a homogeneous spherical system. 

We use the Ricci curvature criterion \cite{GK} of relative instability and perform numerical simulations for static
N-body configurations with given total energy, as a parameter characterizing the system.
It is shown, that in contrast to the role of the central massive body, the core can increase, as well as, decrease
the chaos in the system, depending on the ratio of the core and system's radii. The numerical experiments enable to
obtain the corresponding critical value of the core.

\section{The Method}

The Ricci curvature criterion is defined as follows. Among two systems, the system with smaller negative values of the Ricci curvature in a certain interval $[0, s_*]$ of the affine parameter \cite{GK}
$$
r=\frac{1}{3N}\displaystyle\inf_{0 \leq s \leq s_*} {r_u(s)}.
$$ 
is unstable with higher probability. The curvature 
\begin{equation}
r_u(s) = \frac{R_{ij}u^i u^j}{||u||^2},
\end{equation}
is defined by the Ricci tensor $R_{ij}$ and then is averaged over the directions of the deviation of the close geodesics; $u_i$ is the velocity on the geodesics, for details see \cite{Arn}.

For N-body systems  when the direct impacts of two particles are neglected, which is a fair approximation for real stellar systems, $r_u(s)$ has the form \cite{GK} 
\begin{equation}
r_u(s) = 
-\frac{3N - 2}{2} \frac {W_{i,k} u^i u^k} {W} 
+ \frac{3}{4}(3N - 2) \frac {(W_i u^i)^2} {W^2} 
- \frac{3N-4}{4} \frac {|\bigtriangledown W|^2} {W^3},
\end{equation}
where 
\begin{equation}
W = E - V(r),\,\, u^i = \frac{dr^i}{ds},
\end{equation}
with the derivatives
\begin{equation}
W_i = \frac{\partial{W}}{\partial{x_i}} = \frac{\partial{W}}{\partial{r^k_a}} = -\sum_{c \neq a}{\frac{r^k_{ac}}{r^3_{ac}}},
\end{equation}
when $a \neq b$
\begin{equation}
W_{ij} = \frac{\partial^2{W}}{\partial{x_i}\partial{x_j}} = \frac{\partial^2{W}}{\partial{r^k_a}\partial{r^l_b}} = \left[\frac{\delta_{kl}}{r^3_{ab}} - \frac{3r^k_{ab}r^l_{ab}}{r^5_{ab}}\right],
\end{equation}
when $a = b$
\begin{equation}
W_{ij} = \frac{\partial^2{W}}{\partial{x_i}\partial{x_j}} = \frac{\partial^2{W}}{\partial{r^k_a}\partial{r^l_b}} = -\sum_{c \neq a}{\left[\frac{\delta_{kl}}{r^3_{ac}} - \frac{3r^k_{ac}r^l_{ac}}{r^5_{ac}}\right]}.
\end{equation}
Thus, Eq.(1) defines a measure of relative instability between several systems. 

\section{Analysis}

The created software was generating spherical 3D systems of point particles with random distribution of velocities and coordinates. Equal mass particle systems were considered only, typically with negative total energy i.e. $E_k < |E_p|$ or $e \in (0, 1)$. For given $e$, $k_r$ and $k_m$, the kinetic energy is normalized to 

\begin{equation}
E_k = (1 - e) \left|E_p\right|.
\end{equation}

A sample of $N k_m$ stars is then selected, to generate a subsystem (core), by means of multiplication of the coordinates of its particles by a factor $k_r$; cf. e.g. with the estimations of Ricci curvature in \cite{EZ}. To enable the comparison of the sequence of the modified systems with different sizes and masses of the core, the total energy is initially normalized by means of the variation of the total kinetic energy. This is reached by multiplication of the velocities of all stars by a factor
\begin{equation}
c = \sqrt{\frac{E_k}{E_{k0}}},
\end{equation}
where $E_{k0}$ and $E_k$ are the kinetic energy values before and after the normalization, respectively.

This procedure was repeated for different values of N, starting from $N=250$ up to $8000$. The studied systems revealed similar behavior: they are more unstable at $k_r$ close to both $0$ and $1$, and for $k_m \approx 0$, but not for $k_m \approx 1$. The typical behavior of $r_u$ vs $k_r$ has an extremum, while the dependence on $k_m$ has a plateau for $k_m \approx 1$, as shown in Figure 1.

\section{Discussion}

The Ricci curvature is a local, in time, measure of relative instability, i.e. concerns two or more systems immediately after an initial time moment. The numerical experiments enabled us to study the systems with core radius parameter $k_r$ from $\approx 0$ to $1$, and mass parameter $k_m$ from $0$ to $1$, and total energy parameter $e$ from $\approx 1$ to $\approx 0.5$. Note, the first two parameters are not mutually independent. The typical behavior of the Ricci curvature shows that the core is able both, to increase and decrease the instability of the system, however, its representation via the core parameters $k_r$ and $k_m$ is not equivalent. Namely, when the Ricci curvature is decreasing both at smaller and larger core radii, with maximum value reached at certain critical $k_r^{cr}$, the dependence on $k_m$ is increasing monotonically upon the increase of the latter up to 1.  

Thus, the Ricci curvature criterion enables one to reveal the role of a core in the dynamical properties of spherical gravitating systems by means of relatively simple numerical experiments.

\begin{figure}[thb]
\begin{center}
\psfig{file=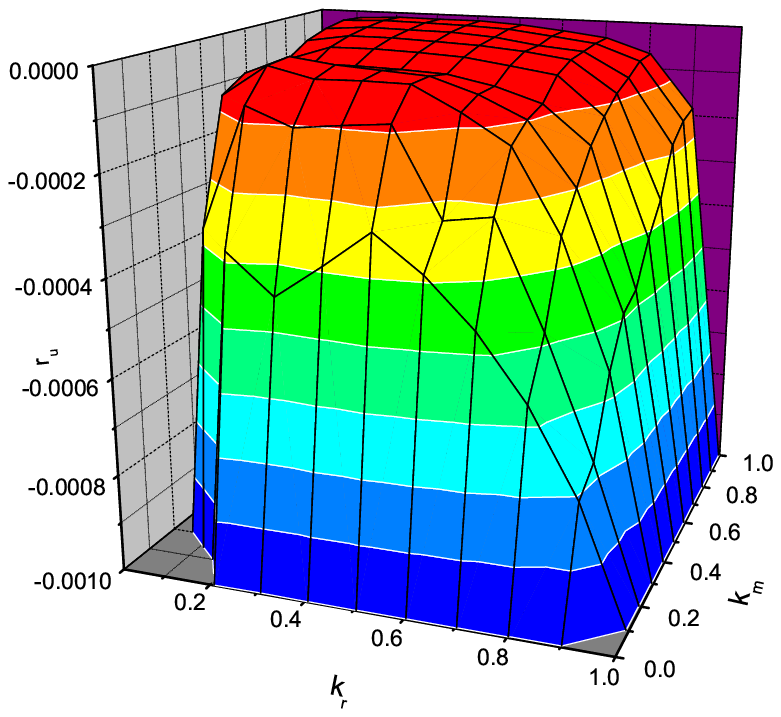,width=5in}
\psfig{file=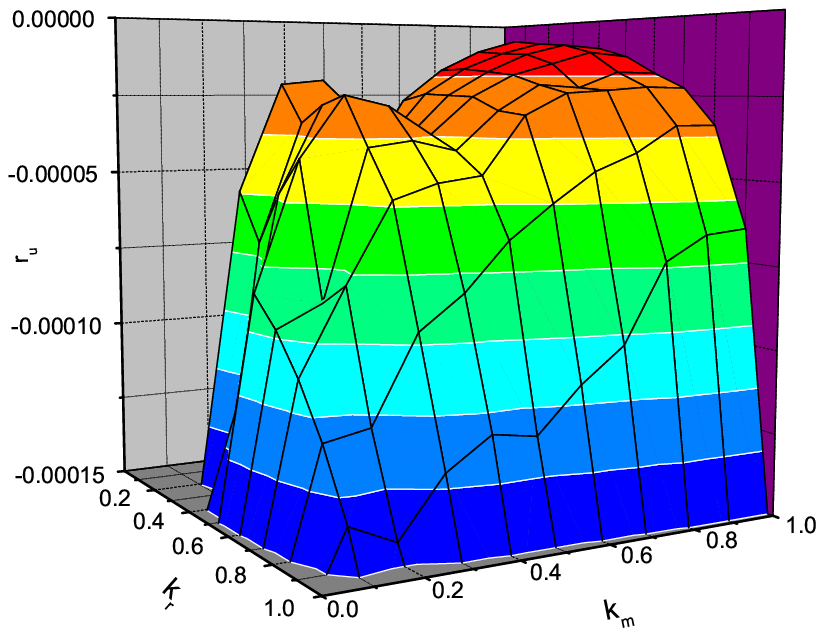,width=5in}
\caption{The dependence of the Ricci curvature on the core parameters $k_r$ and $k_m$, for the normalized energy $e = 0.9$ (top), and $e = 0.81$ (below) for systems of $N=8000$ gravitating particles.}
\end{center}

\begin{center}
\end{center}

\label{aba:fig1}
\end{figure}

\end{document}